# A Longitudinal Study of Student Conceptual Understanding in Electricity and Magnetism

S. J. Pollock

*Department of Physics, University of Colorado, Boulder, CO 80309-0390*

**Abstract.** We have investigated the long-term effect of student-centered instruction at the freshman level on juniors' performance on a conceptual survey of Electricity and Magnetism (E&M). We measured student performance on a research-based conceptual instrument -- the Brief Electricity & Magnetism Assessment (BEMA) -- over a period of 8 semesters (2004-2007). Concurrently, we introduced the University of Washington's *Tutorials in Introductory Physics* as part of our standard freshman curriculum. Freshmen took the BEMA before and after this Tutorial-based introductory course, and juniors took it after completion of their traditional junior-level E&M I and E&M II courses. We find that, on average, individual BEMA scores do not change significantly after completion of the introductory course - neither from the freshman to the junior year, nor from upper-division E&M I to E&M II. However, we find that juniors who had completed a *non*-Tutorial freshman course scored significantly lower on the (post upper-division) BEMA than those who had completed the *reformed* freshman course – indicating a long-term positive impact of freshman Tutorials on conceptual understanding.

**Keywords:** course transformation, course assessment, longitudinal, upper-division electricity and magnetism.
**PACS:** 01.40.-d. 01.40.Fk, 01.40.gb, 01.40.G-

## I. INTRODUCTION AND BACKGROUND

Measurement of the impacts of research-based curricular transformations in introductory physics (through tools such as conceptual surveys) is a primary engine for physics education research [1, 2]. A key question which remains largely unanswered is whether instructional strategies which yield high gains on conceptual tests represent permanent impacts, or if the effects tend to be short-lived. [3,4] With traditional pre-post testing, it is possible that students' performance might represent some form of conceptual rote learning, which would likely disappear over time scales much beyond the examination schedule of individual classes[5,6]. Classic educational psychology literature demonstrates significant loss rates in longitudinal testing even when comparing different initial degrees of learning[7,8], but such studies are frequently clinical studies involving memorized lists rather than conceptual tests. At our institution, we investigated the question of long-term performance on conceptual surveys among physics majors, by tracking a cohort across the multi-year time span between introductory physics and upper-division physics. Several circumstances combined to allow this study. First, we had a distinct and sudden transformation in the structure of our introductory physics II course (Electricity and Magnetism, "E&M") in Fall 2004, namely the introduction of the University of Washington's (UW) *Tutorials in Introductory Physics*[9,10,11]. There is considerable published evidence regarding the efficacy of Tutorials[1,12,13]. Simultaneously, we began to consistently measure conceptual learning gains at the lower division (Physics II) level with a validated research-based instrument, the BEMA (Brief E&M Assessment)[14]. To track our own level of success in implementing Tutorials, we wanted to establish a local baseline of student performance in E&M using the BEMA. Very recently, BEMA data from several peer institutions have been published[15], but when this study began we had no local measurement of freshman BEMA performance in a more traditionally (non-Tutorial) taught E&M course, and so we simultaneously collected data from upper-division physics students in physics "301" and "302" (Junior-level E&M I and II). We initially assumed upper-division physics majors' BEMA performance would provide a reasonable target goal for learning in our large introductory class. We also realized that this would provide an opportunity to measure the *longitudinal* impacts of freshman reforms, roughly four semesters later, by comparing juniors who had been taught without Tutorials (in the early years of the study) to those who had been taught using the Tutorial curriculum (in the later years of the study).

The BEMA is a difficult test - our incoming graduate students average just over 80% correct (N=25 individuals, over 5 years). Recent results from two thousand students in introductory electromagnetism at four institutions[15] confirms our local outcomes - the average BEMA pretest for freshmen is roughly ~25%, and the average BEMA post-test for traditionally taught freshmen is roughly ~45%. For students in our own freshman courses, the correlation coefficient (r) of BEMA post-score to course grade ranges between .45 and .65 for different terms, a fairly high correlation considering the wide variety of components (including homeworks, participation, recitation scores, etc) which go into course grade. This correlation indicates that the BEMA is a useful and relevant measure of learning in our Physics II course, and that it matches with our, often more traditional, choices of assessment.

Preliminary results from the first few years of data [16] showed that juniors' performance on the BEMA was significantly different depending on the students' freshman E&M experience. We have since obtained more upper division data, as well as additional data from our institution[17], allowing for tracking of individuals' course grades across the multi-year period, providing control data on other populations of students taking these instruments (e.g. other majors, or students who took upper-division physics without having taken our introductory sequence.) We have now accumulated *post-test* data for fourteen traditionally taught upper-division E&M I and II courses (Phys 301 and 302), as well as two recent semesters of *pre*-test data for Phys 301, to directly investigate the average decrease of BEMA scores ("fade") between freshman and junior years. We also obtained grade data on the students in these courses (both lower and upper division) who did not take the BEMA in their junior year, to assess our sampling demographics. Our primary results regarding the development of student understanding of E&M concepts, as measured by the BEMA, are:

- We see evidence that the transformation of our introductory courses, including the addition of freshman Tutorials, has had a significant impact (statistically and pedagogically) on student performance on this instrument two years later, at the upper-division level.
- We see evidence that transformed freshman courses with Tutorials have only a small positive impact on overall student performance in junior level classes themselves, as measured by course grades.
- We see evidence for a small drop in performance for individual students over the 2+ year period between when they complete a transformed freshman Physics II and when they start first semester upper-division E&M. (This drop is mostly regained by a small rebound during upper-division E&M.)
- However, we do *not* observe any significant change in performance on the BEMA for individual students between completion of freshman Physics II and completion of (traditionally taught) upper-division E&M I, nor is there any significant shift following the second semester of E&M (E&M II).

Thus, the conceptual learning outcomes of physics majors observed after our transformed freshman physics course persist over time, with no measurable cost at traditional performance at the advanced level, with relatively little impact from the upper-division courses themselves.

## II. COURSES AND STUDENT POPULATIONS

Throughout this paper, "Physics II" refers to the introductory Calculus-based second term course on E&M at CU. The class serves engineers (about 55% of the students), physics/astronomy/engineering physics majors (just under 10% of the students), and a variety of other (mostly) science majors. The course is approximately 25% female. Class size ranges from ~325 to ~475 depending on semester. The basic course structure and content has not changed in many years, and Physics II classes are taught by different faculty every term. For roughly the last decade, almost all faculty at CU have used interactive engagement methods based on *Peer Instruction*[18] in their large lectures. The first use of peer instruction (with colored cards) began in the late 1990's, and clicker technology was introduced in the early 2000's. The courses also consistently use computer-based homework[19], and a staffed help-room for individual instruction. In Fall 2004, traditional recitations were replaced by UW Tutorials[9]. A more detailed description of the transformed classes at CU can be found in our earlier work [10,11,13]. The primary consistent curricular switch which occured in 2004 was the addition of Tutorials with trained undergraduate Learning Assistants [20], in most other respects the course has maintained the same character and syllabus.

"Physics 301" and "Physics 302" are CU's upper-division physics majors' E&M sequence, with an average of ~20-50 students, depending on term (Phys 302 is generally about 20% smaller than 301.) Both courses are offered every term. The vast majority of these students are junior physics, engineering-physics, or astrophysics majors; the class is typically 15% females. The canonical textbook for many years has been Griffiths[21]. The course is taught in a completely traditional physics lecture style by a variety of different faculty (14 faculty have taught these two courses in the last five years) Starting in Spring 2008, we have begun transforming upper-division E&M I[22], so *post-test* data for these semesters are not included in this study[23], but *pre*-test data for the most recent two semesters of Phys 301 are included in our longitudinal discussion.

## III. MEASUREMENTS

The BEMA[14] instrument has been given at the start and end of Physics II every semester since we began using UW Tutorials, issued via paper and scantron during the first and last weeks of the term, providing ten semesters of data. We have obtained matched, valid (e.g. most questions attempted, not all answers the same or other similar patterns) pre/post scores for over 80% of students who complete the course. From Fall 2004 to Spring 2007 the average final Physics II course grade for all students for whom we have matched, valid BEMA pre/post data is 2.80 (on the scale with 0=F, 4=A)($\sigma$=0.84, N=1850). The average course grade over those semesters for *all* students was 2.66 ($\sigma$=.95, N=2242) This *is* a statistically significant difference (p<<.01 level, 2-tailed t-test), due to the large N values, but is not a large pedagogical difference. Nevertheless, this indicates a subtle systematic issue: those students who show up for both the first and last recitation to take the BEMA are on average slightly better students than the small number who do not.

In the upper-division (301 and 302) courses, students were asked only at the *end* of the semester by their instructor to take an online version of the BEMA. This voluntary approach resulted in a slightly smaller fraction of returns than in the freshman course, 66% on average. Overall, it appears the population of students skipping the upper-division post-test is slightly, but not significantly, academically weaker than those who take it, similar to the situation in the lower division. The average course grade in Phys 301 was 3.15 ($\sigma$=.8, N=116) for those who took the BEMA after Phys 301, compared to an average course grade of 2.84 ($\sigma$=.9, N=135) for those who did NOT voluntarily take the BEMA after Phys 301. Here, as at the introductory level, this difference is statistically significant at the p<.01 level (2-tailed t-test), but is perhaps marginally pedagogically significant. For Phys 302, the average course grade of students for whom we have an upper-division BEMA score is 3.03 ($\sigma$=.8, N=119), compared to the remaining students with an average course grade of 2.85 ($\sigma$=.9, N=103, not statistically significantly different with p=.1). Similarly, the average *freshman* grade in Phys II for students who later took the upper-division BEMA is 3.2 ($\sigma$=.65, N=137), compared to 3.0 ($\sigma$=.7, N=68) for those who did not later take the BEMA posttest after either Physics 301 or 302, again a marginally statistically significant difference (p=.03). Overall, it appears the population of students skipping the upper-division post-test is slightly, but not significantly, academically weaker than those who take it, similar to the situation in the lower division.

The BEMA was administered as a *pre*test for Phys 301 starting in Fall 2008, and the collection rate was over 70%. A final question on the online upper-division BEMA asks students to tell us how hard they tried. Over 50% indicated they took it very seriously, and most of the rest indicated they treated it seriously. The tests for the 3% who indicated that they did not take it seriously have been treated as invalid, much as freshman paper results with multiple blanks or simple repeated patterns.

In parallel to our own BEMA data collection, we have obtained course grade data [17] for the ~250 students who have taken Phys 301 or 302 between 2004 and 2007, and all students who took introductory physics dating back to before 2000. We are able to longitudinally follow over 200 individual students from freshman physics through upper-division physics, roughly half of whom took freshman physics at CU *after* Tutorials were implemented. Roughly ~50 students for whom we have upper-division data either took introductory college physics elsewhere, or skipped it due to AP credits.

## IV. DATA AND RESULTS

The Physics II BEMA data were originally collected primarily to evaluate the impact of our curricular reforms, with some results reported elsewhere[10,11,13]. Here we are interested in the longitudinal aspect - an evaluation of differences in results on the BEMA between introductory and upper-division students, as well as the change over time of individual students' scores on this instrument. In the sections that follow, we often characterize/label our transformed freshman course as "with Tutorials", since this was the single most significant and consistent change in our course which characterized our transformation. But we should be clear that a variety of faculty teach these introductory courses, using different materials and with different styles, and thus outcomes measured by the BEMA may be attributable to a number of causes, not just learning from Tutorials. The U. Washington Tutorials are a significant element in our CU transformations, and thus we use this label as a descriptor to distinguish our pedagogical reforms from, e.g. the use of different research-based curricula.

### A. Introductory Physics II

**All students:** The average BEMA *pre*score for all students in Phys II has been very stable for ten terms at 26±1% (with consistent standard deviation, σ~10%). Posttest scores are somewhat more variable, ranging from 50-61%, with an average of 55% overall (σ~16%). The authors of the BEMA[14], and recent outcomes from a cross-institutional study[15] indicate that our prescores are typical, but that our postscores are consistently above what might be expected from a purely traditional introductory lecture course, indicating a level of success of our curricular reforms consistent with outcomes in our reformed Physics I class [10,11]. Our data represent matched, valid BEMA results for over 2600 students[10,13].

**Tracked Students:** Of the large group of introductory students, over 200 individuals went on to later take our upper-division E&M Phys 301 between Fall 2004 and Fall 2007; these are our "tracked students". If we include only this population of tracked students (mostly future physics majors, N=205), we have matched, valid Physics II BEMA data for 2/3 of these, consistent with the collection rate for the course as a whole. The average Phys II BEMA pretest score for these "tracked students" was 33±2%, and their average Phys II BEMA posttest score was 68±2%, both close to a standard deviation above the overall Phys II class averages. Similarly, the average Phys II course grade of these tracked students was 3.2 (on the scale with 0=F, 4=A), more than half a letter grade above the class average. (The average Phys II course grade for students for whom we do not have Phys II BEMA data, but who *did* move on to upper-division physics, was 3.1. Compare these to the average course grade for *all* students, which is 2.66 with σ~1). Thus, the sub-population of Phys II students who will later go on to upper-division physics perform above the freshman class average on all these measures. Any conclusions we draw about long term (longitudinal) impact of introductory reforms should therefore be interpreted as telling us directly only about the sub-population of future physics majors, *not* the introductory-level student population as a whole.

### B. Upper-division Phys 301 and 302.

**Control group:** For the first three terms of this study, none of the upper-division students had gone through an introductory course with Tutorials. Thus, we have accumulated baseline BEMA performance data from upper-division physics majors -- who experienced a mix of non-Tutorial introductory courses -- *after* taking upper-division E&M. We are missing information about Phys II BEMA performance for all students *without* freshman Tutorials, since the BEMA was only given after course reforms were introduced. There are no significant systematic instructor effects when we compare BEMA data or course grades across different semesters of upper-division E&M. (An ANOVA test of BEMA scores for students who have not taken Tutorials, across individual courses, finds no significant difference for any course, with p=.71) The average upper-division BEMA scores for this period were 53±2% (σ =19%) for N=71 unique students[24], representing well over half of all enrolled students. This population was statistically comparable to the rest of the study sample, based on course grades and overall GPA. This is our "control" group, described in the first row of Table I. This average BEMA score is *below* the postscore obtained after our freshman courses with Tutorials, a result we originally found somewhat surprising, and discussed below.

**Bypass group:** Starting in Spring 2006, the upper-division population changed, since most students came up through the usual sequence from a freshman Tutorial experience, while a smaller group did not. This latter group included students who passed out of freshman physics due to AP credit, transfer students, and students who took extra time to go through the sequence. We investigated whether this group (which we refer to as the "Tutorial bypass" group) had different BEMA scores than the control group from earlier terms (who had no opportunity to take Tutorials). Table I summarize the result - the BEMA scores of the "Tutorial bypass" group are higher, but not statistically significantly so, than the control group (p=.10, 2-tailed t-test) Averaging, then, over all terms, upper-division students who do not have a Physics II Tutorial experience score ~56±2% ($\sigma$=20%, N=100) on the BEMA after their junior E&M course. (See also Figure 1, left bar)

Table 1: Summary of upper-division (UD) BEMA scores, and course grades (0-4) over seven semesters of data for both Physics 301 and 302. When scores for 301 and 302 are available, we average them for this table (Course grade in parentheses is calculated for just the subset of students who took the upper-division BEMA).

| Earlier (freshman) experience | Semester time period | UD BEMA average | St. Dev. ($\sigma$) | N (# students with UD BEMA) | UD Course letter grade | $\sigma$ (letter grade) | $N_{tot}$ |
|---|---|---|---|---|---|---|---|
| "control" (No Tutorial) | Fall04-Fall05 | 53±2% | 19% | 71 | 2.9 (3.0) | 0.8 | 111 |
| "bypass" (No Tutorial) | Sp06-Fa07 | 61±4% | 21% | 29 | 2.8 (3.0) | 1.0 | 48 |
| "Tutorial" | Sp06-Fa07 | 71±2% | 15% | 67 | 3.0 (3.24) | 0.8 | 97 |

**Tutorial group:** We now consider the scores for upper-division students who *did* come through freshman Tutorials. The results are shown in the third row of Table 1, and Fig 1. These students (bar on the right of Fig 1a) have an average BEMA score of 71±2%, statistically significantly higher than their non-Tutorial compatriots in the same courses, the "bypass" population (p=.02 for a 2-tailed t-test) and with the control group of *all* upper-division students from earlier terms (p<<.01)[25]. This result is one of the central observations of our study; students who had an introductory freshman Tutorial experience at CU have BEMA scores, when taken *after* their upper-division courses, ~15 points higher than those who didn't. Comparing to data from later semesters for the small group of students taking the *same* upper-division classes who had bypassed Tutorials shows a statistically significant ~10% difference; more indirectly comparing later students to the earlier upper-division control group who never had Tutorials as freshmen we see an 18% difference. These differences are also pedagogically significant; the effect size[26] for UD BEMA scores, comparing Tutorial and "bypass" students in common semesters, is 0.4, and exceeds 0.7 comparing Tutorial students with the control group.

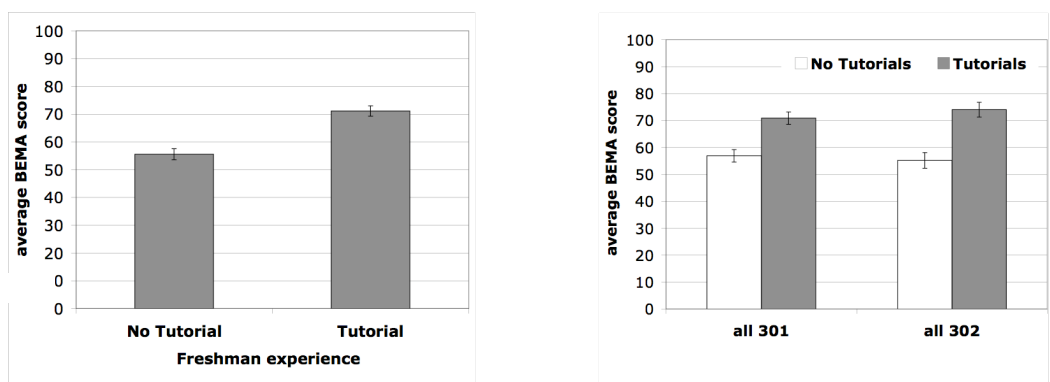

**FIGURE 1. BEMA scores *after* upper-division E&M. Fig a** (left) shows averages over both upper division courses. The left bin shows BEMA scores for students who had never been through a Physics II Tutorial environment. The data are averaged over 7 terms (Fa04-Fa07) and two courses (301 and 302), representing N=100 individuals with BEMA postscores. On the right are those students who had gone through a freshman Tutorial experience. (N=67) The difference (>15%) is significant. **Fig b** (right) separates data for Phys 301 and 302. Errors are standard error of mean.

Does it make a difference whether data are collected after the first or second term of upper division E&M? In Fig 1b, we provide results as above, this time separating Physics 301 and 302. The leftmost bin for each class shows students who never had introductory Tutorials. We see a statistically insignificant difference in BEMA scores from 301 to 302. Fig 1b shows that the difference between Tutorial and non-Tutorial students persists even after 2 semesters of upper-division physics ($p<.01$).

Many faculty have explicit course goals in Phys 301 and 302 which are far from the sort of fundamental conceptual understanding addressed by the BEMA. These courses have an advanced, highly mathematical and quantitative problem-solving focus, which is perhaps more accurately assessed through the course grades than the BEMA. Does the BEMA relate to these more traditional measures of assessment? The correlation coefficient of upper-division BEMA score to upper-division course grade is relatively high (0.57 for N=167 students) But, our upper-division courses are traditionally curved, with the average course grade in both Phys 301 and 302 quite consistent over time (3.0 GPA, "B-centered", with a standard deviation just under 1, for N>250 students over 7 terms. See also the final columns in Table I). For the students in the control group, the average course grade was $2.94\pm0.1$ ($\sigma=0.8$). For students in the "bypass" subgroup in later terms, the average course grade was $2.84\pm0.1$ ($\sigma=1.$), further evidence that this subgroup is not significantly pedagogically different from the earlier control group. For the "Tutorial" subgroup, the average upper-division course grade was $3.04\pm0.08$ (s.d=0.8, N=97), marginally but not statistically significantly higher than students without Tutorials in the *same* upper-division classes. ($p=.20$, 2-tailed t-test) For our faculty colleagues, however, this result has been a very significant outcome - the difference (about 1/5 of a letter grade *higher* for the Tutorial cohort in the same classes) demonstrates that the increased focus on conceptual development at the freshman level is certainly not harming, and likely benefiting, our upper-division majors by our own traditional measures.

## C. Direct Longitudinal Comparisons

The data above indicate long-term benefits of having taken courses which include Tutorials as freshmen, in terms of improved upper-division BEMA scores with no cost in traditionally measured course performance. But does upper-division Phys 301 (or 302) impact students' freshman-level conceptual understanding? Before this study, we had assumed that upper-division E&M courses would move our majors further on the path towards expert-like conceptual performance on the basic topics covered by the BEMA. The outcome was thus something of a surprise; we see little evidence that our traditional upper-division courses positively impact BEMA scores. The overall average BEMA score for *all* students in Phys II in the semesters we're considering was 56%, but the average score for those tracked students who would ultimately take Phys 301 and/or 302 (later in their careers) was close to 70%, so our future majors started well above class averages. As discussed above, we were able to track individuals from introductory level Phys II through upper-division. For the N=38 students for whom we have matched pre-post Freshman *and* post Phys 301 BEMA data, these students gained an average of +39 ($\pm2$) points from pre- to post-Physics II (compared to an average absolute gain of +28 points for the rest of their freshmen classmates), but there was *no* average gain from Post Physics II to Post Phys 301. Figure 2a (left) shows a histogram of the distribution of these shifts. Although some individuals have significantly different scores, the first upper-division E&M course does not have any overall measurable incremental positive impact on conceptual understanding of freshman-level material. We thus observe that our future physics majors are already doing well above class average in their introductory course as freshmen, demonstrate large (and significant) learning gains after introductory physics (significantly higher than the non-physics major colleagues), and retain these skills years later, but the upper-division courses do not have a significant *additional* impact on average student understanding of the basic conceptual issues assessed by the BEMA.

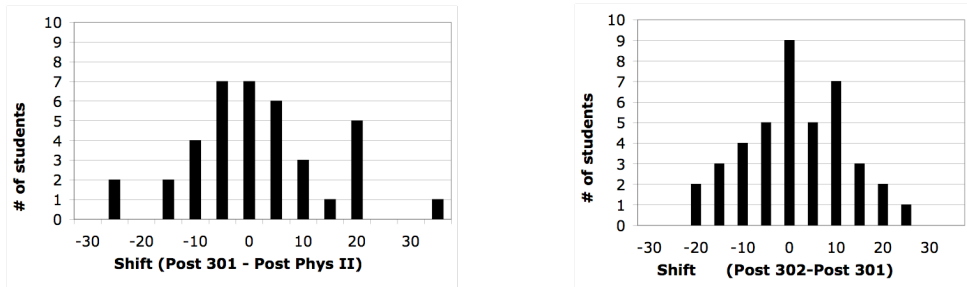

**FIGURE 2. Shift in post-BEMA scores. Fig a (left)** shows a histogram of the shift in individual's scores between freshman and junior courses, (Student score as % after Phys 301) - (Same students' score after Phys II). The average of this distribution of shifts in score is 0 ± 2%, statistically indistinguishable from zero. **Fig b (right)** shows a histogram of the shift in individual's scores across upper-division courses, (student scores after Phys 302) - (same students' score after Phys 301). The average shift is -0.3±2 points (s.d.=11%), statistically indistinguishable from zero.

We can also examine the impact of the second semester of upper-division E&M. Although one might wonder about exam fatigue, 41 students took the BEMA after *both* 301 and 302, and for these students, their average *shift* in BEMA score from 301 to 302 was also indistinguishable from zero. Figure 2b (right) shows the distribution of individual shifts - here again, although some students have significantly shifted scores, the Phys 302 course also does not have any overall measurable incremental positive impact on conceptual understanding of freshman-level material, in aggregate.

## A. Shifts within sub-topics

We have seen several broad results above - better upper-division BEMA performance of students who had freshman Tutorials, with negligible impact of upper division physics on BEMA performance. Do these results arise within specific topical areas, such as those emphasized in specific Tutorials, or do they persist when we look across content areas within the BEMA? We investigate this by examining clusters (3-8 questions per category) of BEMA questions. We also include three additional research-based circuit questions [27]. (The scores for these additional questions have not been included in the data presented earlier, since they are not part of the usual BEMA.) We chose these clusters based on our own ad-hoc evaluation of common broad topical themes. Figure 3 allows a comparison of average scores on these subsets of questions. For each group of questions, we show three averages: black bars show students at the end of Physics II with Tutorials, grey bars show those same students after upper-division Physics, and white bars show the remaining students in upper-division who never had introductory physics with Tutorials. The patterns of differences between these groups, seen earlier for the overall instrument, persists even when zooming in on these narrower subject areas. In each cluster of questions, we see a statistically significant difference in upper division students' scores, depending on whether they did or did not have freshman Tutorials. However, in no category is there a statistically significant change in scores between post-freshman and post-upper division. The largest such shift is in cluster 4, the circuit questions, which is a BEMA topic not revisited in any way during the upper-division course, but even here the decline is not statistically significant (p=.14) Thus, even when looking at different conceptual sub-categories of questions, it appears that the freshman physics experience has a significant positive residual long term impact on future physics majors, but the upper-division courses do not show significant *additional* impacts on student understanding of the basic conceptual issues assessed by the BEMA.

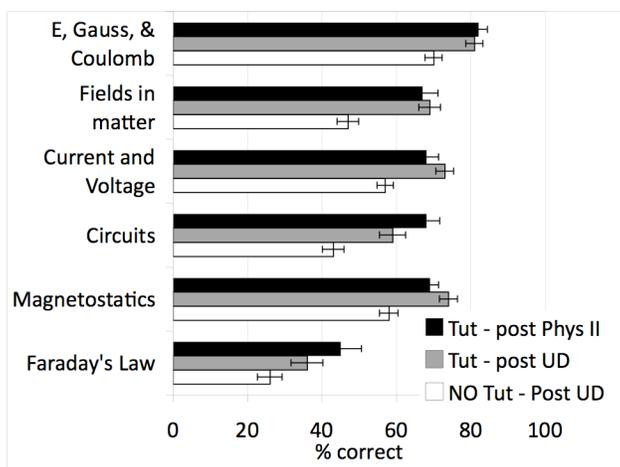

**FIGURE 3. Comparison of performance on sub-groups of BEMA questions.** For each of six clusters of BEMA questions, we show the scores of N=49 individuals for whom we have matched data from Physics II and upper-division (We average post-301 and post-302 scores for individual students when both scores are available) (White bars, consists of N=101 individuals, who had never taken freshman Tutorials. We have no freshman level BEMA data for this population.) Cluster definitions are: E-field, Gauss' law, Coulomb's law (BEMA questions 1-6 and 18). Fields in matter (BEMA 7,12,19). Current and Voltage (BEMA 8,9, 13-16). Circuits (BEMA 10,11,17, and ECCE questions 10-12 [Ref 27]). Magnetostatics (BEMA 20-27, 30). Faraday's Law (BEMA 28, 29, 31)

## B. Upper division pretest study

The lack of a significant shift in posttests from post-Physics II to post-Physics 301 leaves open the question of what students would have scored on the BEMA just before Physics 301. Have students forgotten freshman conceptual material during the intervening years, but then relearned it in the upper-division course? To address this question, for two recent semesters (not included in the post-score results for the data sets above, because the pedagogy of the upper-division course has changed [22,23]), we collected BEMA data at the *start* of the upper-division Phys 301 course. For those students who we could match back to the introductory course (N=38), the BEMA post-physics II score was 61±3%, the pre-Phys 301 score was 56±2%. Thus, for this (new) group of students, there is only a very small, roughly 5% fade over the average 3.6 semesters between introductory and junior level physics. The small magnitude of this fade is somewhat surprising, the rate loss of content knowledge measured in typical education-psychology longitudinal studies [5-8,15] for such a long time scale is typically much higher, but the population studied here is *not* a broad audience, nor is our test primarily factual - these are upper-division physics majors tested on physics concepts. The moderate fade over time for our freshman Tutorial students is statistically significantly different from zero (a matched, *paired* 2-sample t-test yields p=.01) The histogram of the fade is shown in Figure 4, demonstrating the average long-term persistence of conceptual understanding.

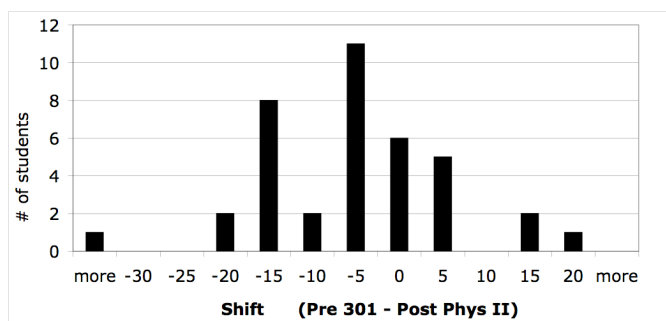

**FIGURE 4. Shift in BEMA scores** between freshman (post) and junior (*pre*). This histogram shows individual student shifts, defined as (student scores just before Phys 301) - (Score from same student just after Phys II). The average shift is -5 ±2%.

Of course, the small size of these shifts cannot be characterized by or attributed solely to introductory pedagogy. The typical physics major's educational path between freshman and junior E&M courses includes several labs, a course on modern physics, and classical mechanics, all of which occasionally touch on topics of electromagnetism. The average BEMA pre-301 score for the N=16 students in these same two semesters who bypassed introductory Physics II, and thus never had Tutorials, was 51±5.5%. This is roughly 5% below the full "freshman Tutorial" population (but not statistically significantly different, p=0.3), shown in Fig. 5. Post 301 data is not shown in this figure, since this population went through a transformed Phys 301 course.[23] For comparison, recent data from students at Carnegie-Mellon [15] shows nearly identical pre-Physics II data. Their transformed Physics II course post scores are higher than ours (~70%), their traditional Physics II post scores are much lower (~53%), and their longitudinal retention study shows roughly 20% drops for *both* populations over time scales of 1 semester to several years. However, their data is for a general population of students, not just future physics majors.

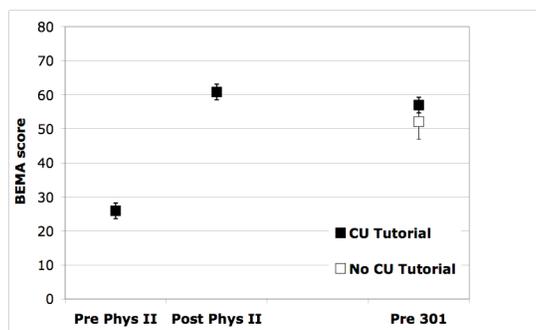

**FIGURE 5. Evolution of BEMA scores** over time. This plot shows the trend of BEMA scores over time for two populations, those who had Tutorials in physics II (N=38), and those who bypassed physics II (N=16), and who took the BEMA as a pretest just at the start of Phys 301. (Of course, we do not have Physics II data for the latter group.)

## V. CONCLUSIONS

By measuring student performance on the BEMA at both freshman and upper-division levels, we are beginning to unpack the impacts of various pedagogies on student conceptual understanding of introductory topics. We find that research-based transformations at the introductory level have produced learning gains for our entire class at levels well above traditional instruction as measured at peer institutions, matching the levels demonstrated by other research-based curricula[15]. Our future physics majors *gain* even more than the average students in those classes. We find by direct measurement that these gains are largely maintained by our majors over the typical 2-year interval between introductory (reformed) classes and the start of upper-division E&M courses, with only a small downwards shift. We also find that our traditional upper-division E&M courses *by themselves* do not appear to have much impact on BEMA scores, evidenced by direct measurement of shifts (or lack thereof) for individuals taking the BEMA before and after either of our upper-division courses. However, students who took our freshman classes with Tutorials still demonstrated enhanced conceptual understanding (and marginally higher course grades) several years after taking their introductory course, with differences after junior E&M typically exceeding 10% on the BEMA above comparable peer upper-division students who never experienced reformed introductory pedagogy. It is interesting to speculate about the origin of these residual differences in the upper-division populations. UW Tutorials do not directly address all of the topics and questions on the BEMA, although there is considerable overlap; nor are Tutorials typically quantitative. Tutorials focus on conceptual understanding, sense-making, and explanations, which appear to manifest in improved performance on this conceptually focused exam. Apparently some of the qualitative understanding built at the introductory level persists over time, and continues to benefit students at the upper-division level, as evidenced by improved BEMA scores and (marginally) improved grades. On the other hand, the common belief that our traditional, perhaps largely technical focus of upper-division courses should consolidate and deepen conceptual understanding of freshman-level concepts is not borne out in our data. We believe such results are of value to PER researchers trying to understand mechanisms and outcomes of reformed curricula such as the Tutorials, and also to traditional faculty trying to decide on the value (and costs) of such curricula.

## ACKNOWLEDGMENTS

Thanks to PhysTEC (APS/AIP/AAPT), NSF CCLI (DUE0410744) and NSF LA-TEST (DRL0554616) for support of our transformed classes. Enormous thanks to Stephanie Chasteen, the CU Physics Department, upper-division faculty and students who helped facilitate this research, the University of Washington's Physics Education Group, and the PER at Colorado group.